\documentclass[12pt,a4paper]{amsart}
\usepackage{amsfonts,amssymb,latexsym,amsmath, amsxtra, eufrak}
\usepackage[all]{xy}
\usepackage{graphics,graphicx,epsfig}
 
 \usepackage{pdfsync}

\usepackage[OT2,T1]{fontenc}
\DeclareSymbolFont{cyrletters}{OT2}{wncyr}{m}{n}
\DeclareMathSymbol{\Sha}{\mathalpha}{cyrletters}{"58}

\theoremstyle{plain}

\newtheorem*{conjecture*}{Conjecture}

\numberwithin{equation}{section}

\let\non\nonumber

\newcommand{\bea}{\begin{eqnarray}}
\newcommand{\eea}{\end{eqnarray}}
\newcommand{\be}{\begin{equation}}
\newcommand{\ee}{\end{equation}}

\newcommand{\sgn}{\mathrm{sgn}}

\newcommand{\noi}{\noindent}

\newcommand{\im}{\mathrm{Im}}

\newcommand{\bP}{\mathbb{P}}
\newcommand{\bZ}{\mathbb{Z}}

\newcommand{\half}{\textstyle{\frac{1}{2}}}

\newcommand{\CC}{\mathcal{C}}
\newcommand{\CH}{\mathcal{H}}
\newcommand{\CM}{\mathcal{M}}
\newcommand{\CN}{\mathcal{N}}
\newcommand{\CO}{\mathcal{O}}

\newcommand{\CZ}{\mathcal{Z}}

\newcommand{\bfk}{\mathbf{k}}




 
\begin{document}

\title[BPS invariants of $\CN=4$ gauge theory on Hirzebruch
surfaces]{BPS invariants of $\CN=4$ gauge theory on Hirzebruch
  surfaces}

\author{Jan Manschot}
\address{Max Planck Institute for Mathematics, Vivatsgasse 7, 53111 Bonn, Germany}
\address{Bethe Center for Theoretical Physics, Bonn Universtity, Nu\ss allee 12, 53115 Bonn, Germany}
\email{manschot@uni-bonn.de}

\begin{abstract}
\baselineskip=18pt
\noi Generating functions of BPS invariants for $\CN=4$ $U(r)$ gauge theory on a 
Hirzebruch surface with $r\leq 3$ are computed. The BPS invariants
provide the Betti numbers of moduli spaces of semi-stable sheaves. 
The generating functions for $r=2$ are expressed in terms of higher level Appell
functions for a certain polarization of the surface. The level corresponds to the self-intersection of the base
curve of the Hirzebruch surface. The non-holomorphic functions are
determined, which added to the holomorphic generating functions provide
functions which transform as a modular form.  
\vspace{.5cm}\\
\end{abstract} 
\maketitle 

\baselineskip=19pt


\section{Introduction}
\setcounter{equation}{0}
The study of supersymmetric spectra of field theories and supergravities
is a major subject in theoretical physics and also mathematics. The
BPS invariant counts the number of BPS states
weighted by a sign. From a more mathematical perspective, the index corresponds to 
topological invariants (e.g. the Euler number or the Betti numbers) of a moduli space
of objects (of an appropriate category) corresponding to the BPS states. 

One of the seminal papers on BPS invariants of
supersymmetric gauge theory on a K\"ahler surface is Ref. \cite{Vafa:1994tf} by Vafa and Witten. They show that the
topologically  twisted path integral localizes on the instanton solutions, and equals
the generating function of the Euler numbers of instanton moduli
spaces, whose natural compactification is the moduli space of semi-stable sheaves. One of their main motivations was to test the strong-weak coupling duality
\cite{Montonen:1977sn} or $S$-duality, which acts by
$SL(2,\mathbb{Z})$ transformations on the theory. 
The coupling constant $g$ and the theta angle $\theta$ combine to the
modular parameter $\tau=\frac{\theta}{2\pi}+\frac{4\pi
  i}{g^2}$. $S$-duality suggests that the generating function of the
BPS invariants (\ref{eq:hrmu}) should exhibit modular properties if the gauge group is  
$SU(r)$ or $U(r)$. They tested this in various cases, for example 
for  sheaves with rank $r=1$ \cite{Gottsche:1990}, and $r=2$ on $\bP^2$
\cite{Yoshioka:1994, Yoshioka:1995, Klyachko:1991}.  The generating
functions for rank 1 were found to be genuine (weakly) holomorphic modular forms.
However the generating functions for rank 2 transform
only approximately as a modular form. These functions are (mixed) mock
modular forms, i.e. functions which do transform as a
modular form only after the addition of  a non-holomorphic
``completion'' \cite{Zwegers:2000}.   

Ref. \cite{Vafa:1994tf} has inspired many results in later years. In
particular, for $r=2$ the dependence of the BPS invariant on the
polarization $J\in H^2(S,\mathbb{Z})$ was included in the generating functions using
indefinite theta functions \cite{Gottsche:1998}. Moreover, the reduced modular
properties for $r\geq 2$ were understood physically as a ``holomorphic
anomaly'' \cite{Minahan:1998vr, Alim:2010cf}.

Although modularity has proven useful for various computations \cite{Minahan:1998vr, Yoshioka:1998ti,
  Gottsche:1998},  physical expectations for $r> 2$ could never be rigorously tested since generating functions for
$r>2$ were not known. 
This was one of the motivations for
\cite{arXiv:1009.1775}, which computed the generating functions of
refined BPS invariants for $r=3$ on $\bP^2$ and its blow-up $\tilde
\bP^2$, which is the Hirzebruch surface $\Sigma_1$.  A convenient
property of $\Sigma_1$ is that the BPS invariants vanish for certain choices
of the first Chern class and choice of polarization. Wall-crossing and the blow-up  formula \cite{Yoshioka:1996}
provide then the invariants in the other chambers and for $\bP^2$.\footnote{Refs. \cite{Kool:2009,
    weist:2009} computed earlier generating functions for the Euler
  numbers for rank 3 using different techniques.}

This article generalizes the computation of the generating function $\CZ_r(z,\rho,\tau;\Sigma_1,J)$ of BPS invariants for $r\leq
3$ of Ref. \cite{arXiv:1009.1775} to more general Hirzebruch surfaces $\Sigma_\ell$, where $-\ell$ is
the self-intersection number of the base curve of $\Sigma_\ell$.  The
arguments $z\in \mathbb{C}$, $\rho\in H^2(\Sigma_\ell,\mathbb{C})$ and $\tau\in \CH$ in $\CZ_r(z,\rho,\tau;\Sigma_\ell,J)$
are generating variables for the Betti numbers of the moduli spaces, and first \& second Chern classes of the
sheaves respectively. 

Section \ref{subsec:rank12} derives expressions for the generating
functions with $r=2$ in terms of indefinite theta functions \cite{Gottsche:1998} and Appell functions of level $\ell$ 
\cite{Appell:1886, Semikhatov:2003uc}. The non-holomorphic but modular
completed functions $\widehat \CZ_2(z,\rho,\tau;\Sigma_\ell,J)$ are
determined for $z\in \mathbb{C}$ (generating function of Betti 
numbers) as well as $z=\half$ (Euler numbers). Due to the presence of these terms
the action of the heat operator $D_r$ on the generating 
function $\widehat \CZ_r(\rho,\tau;\Sigma_\ell,J)$
(\ref{eq:genfunction}) of Euler numbers does not vanish, which is known
in the physics literature as a ``holomorphic anomaly''. A novel
result of the paper is that $D_2\widehat \CZ_2(\rho,\tau;\Sigma_\ell,J)$
in general consists of two terms (\ref{eq:holanomaly}):
\be
D_2\widehat \CZ_2(\rho,\tau;\Sigma_\ell,J)=C_2(\im\, \tau,J) \, \CZ_1(\rho,\tau,\Sigma_\ell)^2+R_2(\rho,\tau;\Sigma_\ell,J),
\ee
where $C_2(\im\, \tau,J)$ is a simple function of $\im\, \tau$ and $J$. The appearance of $\CZ_1(\rho,z,\tau,\Sigma_\ell)^2$ has been
conjectured and discussed in the literature before \cite{Vafa:1994tf,
  Minahan:1998vr, Bringmann:2010sd},
but the additional term $R_2(\rho,z,\tau;\Sigma_\ell,J)$ is novel. 
Remarkably, the additional term vanishes for special choices of
$J$, in particular for $J=-K_\ell$ where $K_\ell$ is the
canonical class of $\Sigma_\ell$. \footnote{Note for $\ell>2$, $-K_\ell$ does not lie
in the ample cone of $\Sigma_\ell$ and is therefore not a permissible
choice for $J$.} 

Another important property of the non-holomorphic completion is that
it  renders $\CZ_r(\rho,z,\tau;\Sigma_\ell,J)$ continuous as a
function of the polarization $J$ \cite{Manschot:2009ia}, which
is expected of a physical path integral.  Although a more intrinsic derivation of the anomaly in physics or
algebraic geometry is desirable, this gives already important insights.
 
Section \ref{subsec:rank3} presents the holomorphic generating
function for $r=3$ (\ref{eq:r3w}) for $r=3$ and presents the Tables 
\ref{tab:betti1}-\ref{tab:betti3}  with the Betti numbers for $\ell=1$. 
The modular properties of $\CZ_3(\rho,z,\tau;\Sigma_\ell,J)$ are 
much more intricate then for $r=2$, and will be discussed elsewhere \cite{Bringmann:2011}.
\\
\newline
The outline of the paper is as follows. Section \ref{sec:sheaves}
reviews the necessary properties of sheaves and Hirzebruch surfaces,
including BPS invariants and their wall-crossing. Section \ref{sec:betti} defines the generating
functions and gives explicit expressions for $r=1, 2$ and 3. 
The non-holomorphic terms and the holomorphic anomaly are determined
for $r=2$ in Subsection \ref{subsec:holanol}, and for $r=3$ Tables with Betti
numbers are presented in \ref{subsec:rank3}.

\section{Sheaves on Hirzebruch surfaces}
\label{sec:sheaves}
\vspace{.2cm}
The Gieseker-Maruyama moduli space of semi-stable sheaves with rank
$r$ on $S$ is the natural compactification of the moduli space of
instantons with gauge group $U(r)$, i.e. anti-self-dual solutions for the field strength: $*F=-F$. The
Chern classes of the sheaf are determined by the topological classes of the instanton:
\be
c_1=\frac{i}{2\pi}\mathrm{Tr}\, F, \qquad
c_2-\frac{1}{2}c_1^2=\frac{1}{8\pi^2}\mathrm{Tr}\, F\wedge F.  \non
\ee
Most of the following is phrased in the more algebraic language of
sheaves, since this setting is most suitable for explicit computations.

\subsection{Sheaves and stability}
The Chern character of a sheaf $F$ on a surface $S$ is given by
ch$(F)=r(F)+c_1(F)+\frac{1}{2}c_1(F)^2-c_2(F)$ in terms of the
rank  $r(F)$ and its Chern classes $c_1(F)$ and $c_2(F)$. The vector
$\Gamma(F):=(r(F),\mathrm{ch}_1(F),\mathrm{ch}_2(F))$ summarizes the
topological properties of $F$. Other
frequently occuring quantities are the determinant
$\Delta(F)=\frac{1}{r(F)}(c_2(F)-\frac{r(F)-1}{2r(F)}c_1(F)^2)$, and 
$\mu(F)=c_1(F)/r(F)$. 

Let $0\subset F_1 \subset F_2\subset \dots \subset F_s=F$ be a
filtration of the sheaf $F$. The quotients are denoted
by $E_i=F_i/F_{i-1}$ with $\Gamma_i=\Gamma(E_i)$. 
With the above notation, the discriminant $\Delta(F)$ is given in terms of the 
topological quantities of $E_i$ and $F_i$ by 
\be
\label{eq:discriminant} 
\Delta(F)=\sum_{i=1}^s\frac{r(E_i)}{r(F)}\Delta(E_i)-\frac{1}{2r(F)}\sum_{i=2}^s
\frac{r(F_{i-1})\,r(F_i)}{r(E_i)} \left(\mu(F_{i-1})-\mu(F_i)
\right)^2. 
\ee
The notion of a moduli space for sheaves is only well defined after
the introduction of a stability condition. To this end let $C(S)\in
H^2(S,\mathbb{Z})$ be the ample cone of $S$.
Given a choice $J\in C(S)$, a sheaf $F$ is called $\mu$-stable if for
every subsheaf $F'\subset F$,
$\mu(F')\cdot J <\mu(F)\cdot J$, and $\mu$-semi-stable if $\mu(F')\cdot J \leq\mu(F)\cdot J$. A wall of marginal stability $W$ is a (codimension
1) subspace of $C(S)$, such that $(\mu(F')-\mu(F))\cdot J=0$, but
$(\mu(F')-\mu(F))\cdot J\neq 0$ away from $W$. 

Let $S$ be a K\"ahler surface, whose intersection pairing on
$H^2(S,\mathbb{Z})$ has signature $(1,b_2-1)$.  Since at a wall,
$(\mu_2-\mu_1)\cdot J=0$ and $J^2>0$, we have $(\mu_2-\mu_1)^2<0$. Therefore, the set of semi-stable filtrations for $F$, with
$\Delta_i\geq 0$ is finite. The ample class $J$
provides natural projections $\mathbf{c}_\pm$ for an element
$\mathbf{c}\in H^2(S,\mathbb{Z})$ to the positive and negative definite
subspaces of $H^2(S,\mathbb{R})$:
\be
\label{eq:projections} 
\mathbf{c}_+=\frac{\mathbf{c}\cdot J\,J}{J^2}, \qquad \mathbf{c}_-=\mathbf{c}-\mathbf{c}_+.
\ee

\subsection{Some properties of ruled surfaces}
\label{sec:ruled}
A ruled surface is a surface $\Sigma$ together with a surjective
morphism $\pi: \Sigma\to C$ to a curve $C$, such that the fibre $\Sigma_y$
is isomorphic to $\mathbb{P}^1$ for every point $y\in C$. Let $f$ be the
fibre of $\pi$, then
$H_2(\Sigma,\mathbb{Z})=\mathbb{Z}C\oplus\mathbb{Z}f$, with
intersection numbers $C^2=-\ell<0$, $f^2=0$ and $C\cdot f=1$. The canonical class is
$K_\Sigma=-2C+(2g-2-\ell)f$. The holomorphic Euler characteristic
$\chi(\CO_\Sigma)$ is for a ruled surface $1-g$. An ample class
 is parametrized by $J_{m,n}=m(C+\ell f)+nf\in C(\Sigma)$ with $m,n>0$. 
The following only considers surfaces with $g=0$, these are known as
rationally ruled surfaces or Hirzebruch surfaces. They are denoted by
$\Sigma_\ell$ and furthermore $K_\ell$ denotes the canonical class.

To learn about the set of semi-stable sheaves on $\Sigma_\ell$ for $J\in C(S)$, it is
useful to first consider the restriction of the sheaves on $\Sigma_\ell$
to $f$. Namely the restriction to $E_{|f}$ is stable if and only if
$E$ is $\mu$-stable for $J=J_{0,1}$ and in the adjacent chamber
\cite{Huybrechts:1996}. However, since every bundle of rank $\geq 2$ on
$\mathbb{P}^1$ is a sum of line bundles,
there are no stable bundles with $r\geq 2$ on $\bP^1$. Therefore, the BPS invariant
$\Omega(\Gamma,w;J)$ (defined in the next subsection) vanishes for
$\Gamma=(r(F),-C-\alpha f,\mathrm{ch}_2)$ with $r(F)\geq 2$ and   
$\alpha=0,1$. 


\subsection{Invariants and wall-crossing}
The moduli space $\CM_J(\Gamma)$ of semi-stable sheaves (with respect
to the ample class $J$) whose rank and Chern classes are determined by $\Gamma$ has
complex dimension: 
\be
\label{eq:dimension}
\dim_{\mathbb{C}} \CM_J(\Gamma)=2r^2\Delta-r^2\chi(\CO_S)+1.
\ee

To define the refined BPS invariants $\Omega(\Gamma,w;J)$ in
an informal way, let
$p(X,s)=\sum_{i=0}^{2\dim_\mathbb{C}(X)}b_is^i$, with $b_i$
the Betti numbers $b_i=\dim H^2(X,\mathbb{Z})$, be the Poincar\'e
polynomial of a compact complex manifold $X$. Then:
\be
\label{eq:refw}
\Omega(\Gamma,w;J):=\frac{w^{-\dim_\mathbb{C}\CM_J(\Gamma)}}{w-w^{-1}}\,
p(\CM_J(\Gamma),w). 
\ee
The rational refined invariants are defined by \cite{arXiv:1009.1775}:
\be
\label{eq:rational}
\bar \Omega(\Gamma,w;J)=\sum_{m|\Gamma}
\frac{\Omega(\Gamma/m,-(-w)^m;J)}{m}. \non 
\ee
See \cite{Manschot:2010qz} for a physical motivation of these rational
invariants and \cite{Kontsevich:2008, Nakajima:2007} for mathematical
motivations. The numerical BPS invariant $\Omega(\Gamma;J)$ follows
from the $\Omega(\Gamma,w;J)$ by:
\be
\label{eq:owtoo}
\Omega(\Gamma;J)=\lim_{w\to -1}
(w-w^{-1})\,\Omega(\Gamma,w;J),
\ee
and similarly for the rational invariants $\bar \Omega(\Gamma;J)$.

A crucial tool for the computation of the generating functions in
Section \ref{sec:betti} is the
wall-crossing formula, which provides the
change $\Delta\Omega(\Gamma;J_\CC\to J_{\CC'})$ across walls of marginal
stability. Ref. \cite{Yoshioka:1995} gives as criterion for his
wall-crossing formula for $r=2$ that $K_\ell\cdot J<0$, which holds
for any $\ell$ and $J\in C(\Sigma_\ell)$. For $r=3$ more complicated
wall-crossings appear, in particular walls where the slope of three
rank 1 sheaves with different $c_1$ become equal. Physical arguments
suggest that for these walls one could use the wall-crossing formulas
of Kontsevich-Soibelman \cite{Kontsevich:2008} or
Joyce-Song \cite{Joyce:2008} since they are shown to hold in both
supergravity and field theory \cite{Denef:2007vg, Andriyash:2010qv,
  Gaiotto:2008cd}. These wall-crossing formulas are derived for  
Donaldson-Thomas invariants, which are defined for 6-dimensional
gauge theory on a Calabi-Yau 3-fold \cite{Donaldson:1998}.
The mathematical justification for the use of these
wall-crossing formulas for sheaves on surfaces is therefore 
not well established. Ref. \cite{Joyce:2004} gives as criterion for
the applicablity that $K_S^{-1}$ must be numerically effective
(i.e. $-K_S\cdot D\geq 0$ for any curve in
$D\in H^2(S,\mathbb{Z})$). This would exclude the Hirzebruch surfaces
with $\ell>2$. The generating function (\ref{eq:r3w}) for $r=3$
is consistent with the wall-crossing formulas for DT-invariants and in
agreement with previous results in the literature for $\ell=1$,  but
in view of the above requires at least for $\ell>2$ further
justification.

Keeping in mind these comments, I continue by giving the
explicit change of the invariants in case of primitive wall-crossing. 
To this end, define the following quantities: 
\be
\label{eq:intprod} 
\left<\Gamma_1,\Gamma_2\right>=r_1r_2(\mu_2-\mu_1)\cdot K_S , \qquad \mathcal{I}(\Gamma_1,\Gamma_2;J)=r_1r_2(\mu_2-\mu_1)\cdot J.\non
\ee
The change $\Delta\Omega(\Gamma_1+\Gamma_2,w;J_\CC\to J_{\CC'})$ for
$\Gamma_1$ and $\Gamma_2$ primitive is \cite{Yoshioka:1996, Kontsevich:2008}
\begin{eqnarray}
\label{eq:DOm}
\Delta \Omega(\Gamma,w;J_\CC\to J_{\CC'})&=&-\half \left(
  \sgn(\mathcal{I}(\Gamma_1,\Gamma_2;J_{\CC'}))-
  \sgn(\mathcal{I}(\Gamma_1,\Gamma_2;J_\CC)) \right) \\
&&\times\left( w^{\left<\Gamma_1,\Gamma_2\right>}- w^{-\left<\Gamma_1,\Gamma_2\right>}\right)\Omega(\Gamma_1,w;J)\,\Omega(\Gamma_2,w;J).\non
\end{eqnarray}
with 
$$
\sgn(x)=\left\{ \begin{array}{cl} 1, & x>0, \\ 0, & x=0, \\ -1, & x<0.\end{array} \right.
$$
The subscript ${W_\CC}$ in $J_{W_\CC}$ refers to a point in $\CC$
which is sufficiently close to the wall $W$, such that no wall is
crossed for the constituent between the wall and $J_{W_\CC}$. Note
that the wall is independent of $c_2$.   

For the computation of the invariants of rank 3, one also needs to
determine the wall-crossing formula across walls of marginal
stability for non-primitive charges $2\Gamma_1+\Gamma_2$ and walls
where the slope of three non-parallel charges becomes equal. These can
be determined using the wall-crossing formulas \cite{Kontsevich:2008,
Joyce:2008}. The result takes a 
simple form in terms of rational invariants and (\ref{eq:DOm})
\cite{Manschot:2010xp}.

\section{Generating functions}
\label{sec:betti}

This section computes the generating functions of the BPS invariants
$\Omega(\Gamma,w; J)$. We start by defining the generating
functions and a brief discussion of their properties. The generating
function $\mathcal{Z}_r(\rho,z,\tau;S,J)$ for a K\"ahler surface $S$ is defined by:
\begin{eqnarray}
\label{eq:genfunction}
\mathcal{Z}_r(\rho,z,\tau;S,J)&=&\sum_{c_1, c_2}\,\bar \Omega(\Gamma,w;J)\,(-1)^{rc_1\cdot K_S}\,\non\\
&&\times \bar q^{r \Delta(\Gamma) -\frac{r\chi(S)}{24}-\frac{1}{2r}(c_1+rK_S/2)^2_-} q^{
  \frac{1}{2r}(c_1+rK_S/2)^2_+} e^{2\pi i \rho \cdot (c_1+rK_S/2)},\non
\end{eqnarray}
with $\rho \in H^2(S,\mathbb{C})$, $w=e^{2\pi i z}$ and $q=e^{2\pi i
  \tau}$. 
Twisting by a line bundle leads to an isomorphism of moduli
spaces. It is therefore sufficient to determine $\Omega(\Gamma,w;J)$
only for $c_1 \mod r$, and it moreover implies that $\mathcal{Z}_r(\rho,z,\tau;S,J)$ allows a
theta function decomposition: 
\be
\label{eq:thetadecomp}
\mathcal{Z}_r(\rho,z,\tau;S,J)=\sum_{\mu\in \Lambda^*/\Lambda} \overline{h_{r,\mu}(z,\tau;S,J)}\,\Theta_{r,\mu}(\rho,\tau;S),
\ee
where the bar over $h_{r,\mu}(z,\tau;S,J)$ denotes complex
conjugation, and $h_{r,\mu}(z,\tau;S,J)$ and
$\Theta_{r,\mu}(\rho,\tau;S)$ are defined by:
\begin{eqnarray}
\label{eq:hrmu}
h_{r,\mu}(z,\tau;S,J)&=&\sum_{c_2} \bar \Omega(\Gamma,w;J)\,q^{r\Delta(\Gamma)-\frac{r\chi(S)}{24}},\\
\Theta_{r,\mu}(\rho,\tau;S)\,\,\,\,\, &=&\sum_{\bfk \in H^2(S,r\mathbb{Z}) +rK_S/2+\mu} (-1)^{r\bfk
  \cdot K_S} q^{\bfk^2_+/2r}\bar q^{-\bfk^2_-/2r} e^{2\pi i\rho\cdot 
  \bfk } .\non 
\end{eqnarray}
Note that $\Theta_{r,\mu}(\rho,\tau;S)$ depends on $J$ through
$\bfk_\pm$ and does not depend on $z$.  

The generating function of the numerical invariants $\Omega(\Gamma;J)$ follows simply from Eq. (\ref{eq:owtoo}):
\be
\mathcal{Z}_{r}(\rho,\tau;S,J)=\lim_{z\to \frac{1}{2}} \quad
(w-w^{-1})\, \mathcal{Z}_{r}(z,\rho,\tau;S,J). \non 
\ee
Physical arguments imply that this function transforms as a
multivariable Jacobi form of weight $(\frac{1}{2},-\frac{3}{2})$
\cite{Vafa:1994tf, Manschot:2008zb} with a non-trivial multiplier system. For
rank $>1$ this is only correct after the addition of a suitable non-holomorphic
term \cite{Vafa:1994tf, Minahan:1998vr}. This is explained for $r=2$
in  Subsections \ref{subsec:rank12} and \ref{subsec:holanol}.

The functions $h_{r,c_1}(z,\tau)$ and $h_{r,c_1}(\tau)$ contain a
factor which depends only on the rank $r$ and $b_2(S)$. 
It is therefore useful to define 
\begin{eqnarray}
\label{eq:fvh}
f_{r,c_1}(z,\tau)&=&\left(\frac{i}{\theta_1(2z,\tau)\eta(\tau)^{b_2(S)-1}}\right)^{-r}
h_{r,c_1}(z,\tau),\non \\
f_{r,c_1}(\tau)&=&\left(\frac{1}{\eta(\tau)^{\chi(S)}}\right)^{-r}
h_{r,c_1}(\tau),\non
\end{eqnarray}
with $\theta_1(z,\tau)$ and $\eta(\tau)$ defined by (\ref{eq:etatheta}).
The function $f_{r,c_1}(\tau)$ follows from $f_{r,c_1}(z,\tau)$ by
\be
\label{eq:fwtof}
f_{r,c_1}(\tau)=\frac{(-1)^{r-1}}{2^{r-1}(r-1)!}\,\frac{1}{(2\pi 
  i)^{r-1}} \partial_z^{r-1} f_{r,c_1}(z,\tau)\vert_{z=\frac{1}{2}}.  
\ee
Note that the terms of degree $<r-1$ in the Taylor expansion with
respect to $z$ of $f_{r,\mu}(z,\tau)$ vanish. 

A useful relation is the ``blow-up formula'' which relates the
generating function of a surface $S$ with that of its blow-up $\phi: \tilde S \to S$ at a non-singular point. Let $C_1$ be the exceptional
divisor of $\phi$, and take $J\in C(S)$, $r$, and $c_1$ such that
$\gcd(r,c_1\cdot J)=1$. The generating functions $h_{r,c_1}(z,\tau;S,J)$ and
$h_{r,c_1}(z,\tau;\tilde S,J)$ are then related by
\cite{Yoshioka:1994, Vafa:1994tf, Yoshioka:1996, Li:1999, Gottsche:1998}:
\be
\label{eq:blowup}
h_{r,\phi^* c_1-kC_1}(z,\tau;\tilde S, J)=B_{r,k}(z,\tau)\, h_{r,c_1}(z,\tau;S,J),
\ee
with
\be
\label{eq:Brk}
B_{r,k}(z,\tau)=\frac{1}{\eta(\tau)^r}\sum_{\sum_{i=1}^ra_i=0 \atop a_i\in \mathbb{Z}+\frac{k}{r}}
q^{\frac{1}{2}\sum_{i=1}^ra_i^2} w^{\sum_{i<j}a_i-a_j}.  \non 
\ee

\subsection{Rank 1 and 2}
\label{subsec:rank12}
This subsection presents explicit expressions for
$h_{r,c_1}(z,\tau;\Sigma_\ell,J_{m,n})$. The result for $r=1$ and
$S=\Sigma_\ell$ is simply \cite{Gottsche:1990}: 
\be
f_{1,c_1}(z,\tau;\Sigma_\ell)=1.\non
\ee
Note that the dependence on $J$ could be omitted here since all rank 1
sheaves are stable. Moreover, there is also no dependence on $\ell$.

\begin{figure}[h!]
\centering
\includegraphics[totalheight=7cm]{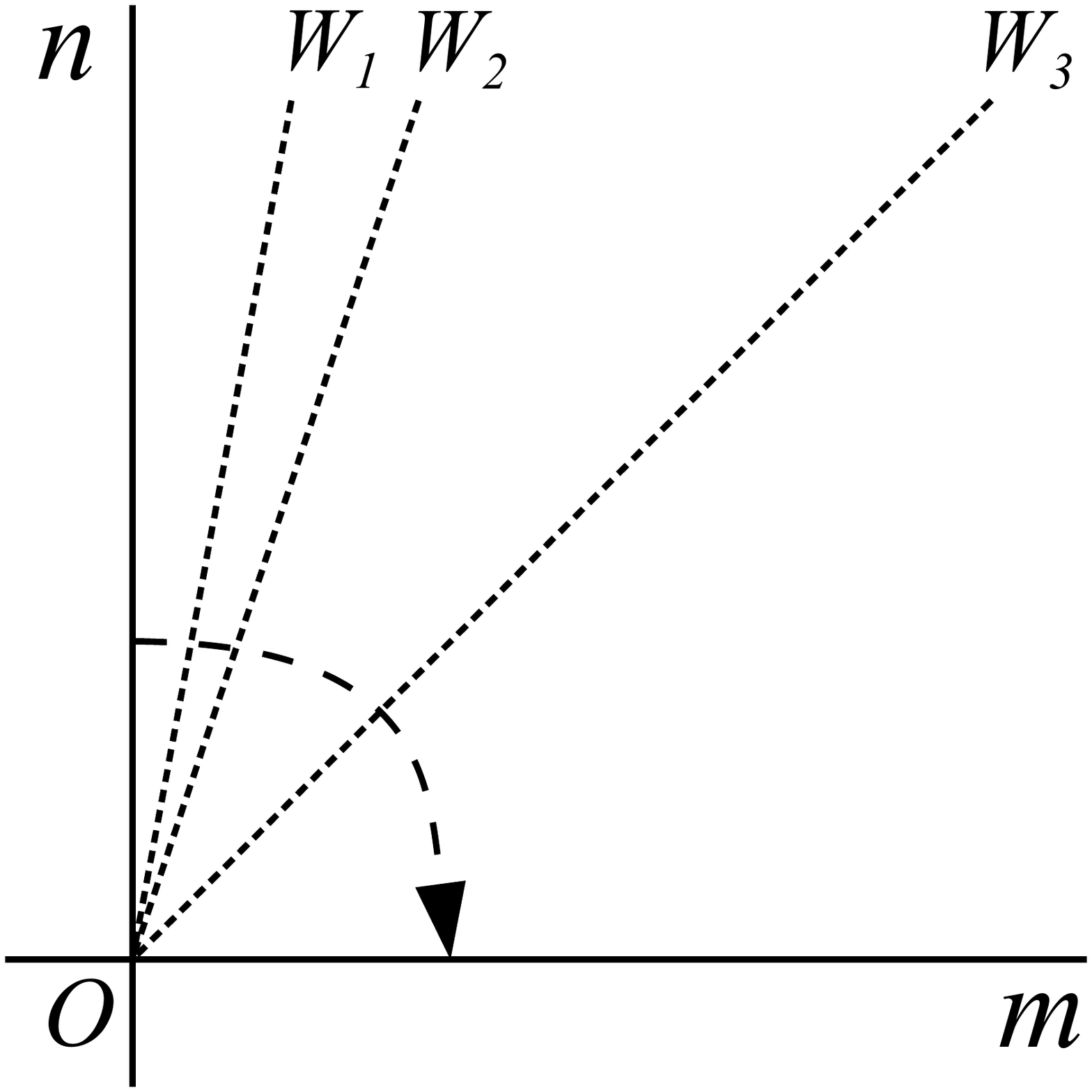} 
\caption{The ample cone of $\Sigma_1$, together with the
  three walls for $\Gamma=(2,-C-f,2)$, namely for $(a,b)=(1,0)$,
  $(2,0)$, $(3,0)$.} 
\label{fig:mspace}
\end{figure}

To compute the generating functions for $r\geq 2$, we use 
wall-crossing together with the fact that
$\Omega(\Gamma,w;J_{0,1})=0$ for $c_1=-C+\alpha f$ and $r\geq 2$.
In the following, $c_1(E_2)$ is parametrised by $bC-af$. The walls are then at
$\frac{m}{n}=\frac{2b-\beta}{2a-\alpha}$ for $r=2$, with $m,n\geq 0$.  
See Figure \ref{fig:mspace} for the walls for $\Delta(F)=\frac{9}{4}$, $r(F)=2$.
One finds \cite{Gottsche:1998, arXiv:1009.1775} using Eq. (\ref{eq:discriminant}):
\begin{eqnarray}
\label{eq:genf2w}
f_{2,C-\alpha f}(z,\tau;\Sigma_\ell,J_{m,n})&=&-\half \sum_{a,b\in
  \mathbb{Z}} \half (\,\sgn((2b+1)n-(2a-\alpha)m)-\sgn(2b+1)\,)\non\\
&&\times \left(w^{(\ell-2)(2b+1)+2(2a-\alpha)}-w^{-(\ell-2)
    (2b+1)-2(2a-\alpha)} \right)\\
&&\times\, q^{\frac{\ell}{4}
  (2b+1)^2+\frac{1}{2}(2b+1)(2a-\alpha)}.\non
\end{eqnarray}
These functions are indefinite theta functions \cite{Gottsche:1996},
which are sums over a subset of the positive definite sublattice of an
indefinite lattice. Since the sum is only over a subset of the
lattice, they transform as a modular form only after addition of a suitable 
non-holomorphic term (depending on $\bar \tau$ and $\bar z$)
\cite{Zwegers:2000}. 

The computation of the invariants for $c_1=-\alpha f$ is much more
involved since strictly semi-stable sheaves do exist for $J_{0,1}$ or if
$\alpha=0$ for every $J\in C(\tilde \bP^2)$. We will circumvent this computation
by determining the functions $f_{2,-\alpha f}(z,\tau;\Sigma_\ell,J_{1,0})$ from modular transformations of
$f_{2,C-\alpha f}(z,\tau;\Sigma_\ell,J_{1,0})$. One can consequently
determine the invariants for arbitrary $J_{m,n}$ by application of the wall-crossing formula. 

We continue by writing $f_{2,C-\alpha f}(z,\tau;\Sigma_\ell,J_{m,n})$ in
terms of two new functions $A_{\ell,(\alpha,\beta)}(z,\tau)$ and $\vartheta_{\alpha,
  \beta}^{m,n}(z,\tau)$: 
\begin{eqnarray} 
&&f_{2,\beta C -\alpha f}(z,\tau;\Sigma_\ell,J_{m,n})  =
A_{\ell,(\alpha,\beta)}(z,\tau)+\vartheta_{\alpha, \beta}^{m,n}(z,\tau),\qquad
\alpha,\beta \in\{0,1\}.\non
\end{eqnarray}
with 
\begin{eqnarray}
\label{eq:indeftheta}
\vartheta_{\alpha, \beta}^{m,n}(z,\tau)&=&\sum_{a,b \in \mathbb{Z}}\half
\left(\sgn(-(2a-\alpha))-\sgn((2b-\beta)n-(2a-\alpha)m) \right)\non\\
&&\times w^{(\ell-2)(2b-\beta)+2(2a-\alpha)}\,q^{\frac{\ell}{4}(2b-\beta)^2+\frac{1}{2}(2b-\beta)(2a-\alpha)}.\non
\end{eqnarray}
Then Eq. (\ref{eq:genf2w}) gives for $A_{\ell,C-\alpha f}(z,\tau)$ for
$\ell\geq 1$ after performing a geometric sum:\footnote{Note that for
  $\Sigma_{\ell=0}=\bP^1\otimes \bP^1$, the
  function $A_{0,(0,1)}(z,\tau)$ is undefined while $A_{0,(1,1)}(z,\tau)=0$.}
\begin{eqnarray}
\label{eq:Al}
&&A_{\ell,(1,1)}(z,\tau)=q^{\frac{\ell+2}{4}}w^{\ell}\sum_{n\in
\mathbb{Z}}\frac{q^{\ell n(n+1)+n}w^{2(\ell-2)n}}{1-q^{2n+1}w^4}, \\ 
&&A_{\ell,(0,1)}(z,\tau)\quad\, =-\frac{1}{2}\sum_{n\in \mathbb{Z}}q^{\frac{\ell}{4} (2n+1)^2}w^{(\ell-2)(2n+1)}+q^{\frac{\ell}{4}}w^{\ell-2}\sum_{n\in
\mathbb{Z}}\frac{q^{\ell n(n+1)}w^{2(\ell-2)n}}{1-q^{2n+1}w^4}.\non
\end{eqnarray}

The functions in Eq. (\ref{eq:Al}) are
specializations of higher level Appell   
functions \cite{Appell:1886, Zwegers:2010}, whose definition is recalled in Appendix A.
These functions appeared earlier in mathematical physics in the theory of characters of superconformal
algebras \cite{Eguchi:1988af, Kac:2000, Semikhatov:2003uc}. See
\cite{Troost:2010ud} for a recent discussion. This might not be
accidental since $\CN=4$ Yang-Mills is well known to be related
related to 2d conformal field theory by M-theory
\cite{Minahan:1998vr}. Deriving these functions explicitly from a
2-dimensional perspective is an interesting direction for future 
research.  

Analogously to the indefinite theta functions, the Appell functions
only transform as a modular (or Jacobi) form after addition
of a non-holomorphic term. Eq. (\ref{eq:appellcom}) gives the exact
expression obtained by \cite{Zwegers:2010}.  Application of this to
our case of interest gives for the completion $\widehat A_{\ell,(\alpha,\beta=1)}(z,\tau)$
:  
\begin{eqnarray}
\label{eq:Appellcomp2}
\widehat
A_{\ell,(\alpha,\beta)}(z,\tau)&=&A_{\ell,(\alpha,\beta)}(z,\tau)+\frac{1}{2}\sum_{k=0}^{\ell-1}\left(
  \sum_{ 
    n_1=2k+\beta \ell+\alpha \atop \mod 
    2\ell}w^{\frac{\ell-2}{\ell}n_1}q^{\frac{n_1^2}{4\ell}} \right)
\non \\
&&\times \sum_{n_2=-2k-\alpha \atop \mod 2\ell}\left(
  \sgn(n_2)-E\left(\left(n_2+2(\ell+2)\,\im\, z/y\right)\sqrt{y/\ell}\right)\right)
w^{-\frac{\ell+2}{\ell}n_2}q^{-\frac{n_2^2}{4\ell}},
\end{eqnarray} 
with $y=\im\, \tau$ and $E(x)=2\int_0^x e^{-\pi u^2}du$.
The four functions $\widehat A_{\ell,(\alpha,\beta)}$ transform as a
vector-valued Jacobi form of weight 1 and index $-8$ of
$SL(2,\mathbb{Z})$ \cite{Semikhatov:2003uc, Zwegers:2010}. One finds
for the action of the generators $S$ and $T$ : 
\begin{eqnarray}
\label{eq:modtrA}
&S:&\quad \widehat A_{\ell,(\alpha,\beta)} \left(\frac{z}{\tau},\frac{-1}{\tau}\right)=\frac{\tau}{2}
e^{2\pi i(- \frac{8 \, z^2}{\tau})} \sum_{\tilde \alpha,\tilde \beta \in \{0,1\}}  (-1)^{\ell
  \beta\tilde \beta+\alpha\tilde \beta+\beta \tilde \alpha} \widehat A_{\ell,(\tilde \alpha,\tilde \beta)}(z, \tau ),\\
&T:&\quad \widehat A_{\ell,(\alpha,\beta)}
\left(z,\tau+1\right)\,= e^{2\pi i \frac{\beta^2+2\alpha\beta}{4}}
\widehat A_{\ell,(\alpha,\beta)} \left(z,\tau\right)\non 
\end{eqnarray}

The modular transformations (\ref{eq:modtrA})
together with the single pole in $z$ of the refined invariants (\ref{eq:refw}) do fix the functions $A_{\ell,(\alpha,0)}(z,\tau)$ to be:
\begin{eqnarray}
\label{eq:Al10}
&&A_{\ell,(1,0)}(z,\tau)=w^2\sum_{n\in\mathbb{Z}}\frac{q^{\ell
    n^2+n}w^{2(\ell-2)n}}{1-q^{2n}w^4}+\frac{i\,\eta(\tau)^3}{\theta_1(4z,\tau)},\non\\ 
&&A_{\ell,(0,0)}(z,\tau)\quad \, =-\frac{1}{2}\sum_{n\in \mathbb{Z}}q^{\ell
  n^2}w^{2(\ell-2)n}+\sum_{n\in\mathbb{Z}}\frac{q^{\ell
    n^2}w^{2(\ell-2)n}}{1-q^{2n}w^4}+\frac{i\,\eta(\tau)^3}{\theta_1(4z,\tau)}. \non 
\end{eqnarray}
This agrees for $c_1=f$ with the generating function in Ref. \cite{Yoshioka:1995}
(Corollary 3.4). The completion of these functions is given by
Eq. (\ref{eq:Appellcomp2}). 

One can show the following relation between $A_{1,(\alpha,0)}(z,\tau)$
and $A_{1,(\alpha,1)}(z,\tau)$ using the quasi-periodicity formula
(\ref{eq:quasiperiod}):   
\begin{eqnarray}
\label{eq:multiplicative}
&&A_{1,(1,0)}(z,\tau)  = \frac{\theta_2(2z,2\tau)}{\theta_3(2z,2\tau)} \,A_{1,(1,1)}(z,\tau), \\
&&A_{1,(0,0)}(z,\tau) \,\,\,\,=
\frac{\theta_3(2z,2\tau)}{\theta_2(2z,2\tau)}\,A_{1,(0,1)}(z,\tau). \non
\end{eqnarray}
This relation is understood in algebraic geometry by the blow-up  formula
(\ref{eq:blowup}), which relates the functions $h_{2,c_1}(z,\tau;\Sigma_1,J_{1,0})$ with
$c_1=C-\alpha f$ to those with $c_1=-\alpha f$. For
$h_{2,c_1}(z,\tau;\bP^2)$ one recovers the result of \cite{Bringmann:2010sd}. 
The multiplicative relation (\ref{eq:multiplicative}) does not hold for $\ell>1$, since $\Sigma_{\ell>1}$ is the blow-up of the weighted
  projective plane $(1,1,\ell)$ at its {\it singular} point  
  \cite{Dolgachev:1982}, and the blow-up formula is thus not applicable.

What remains is to complete the indefinite theta functions
$\vartheta_{\alpha, \beta}^{m,n}(z,\tau)$. One finds using
Ref. \cite{Zwegers:2000}: 
\begin{eqnarray}
\label{eq:wvarth}
\widehat{\vartheta}_{\alpha, \beta}^{m,n}(z,\tau)&=&\sum_{a,b \in \mathbb{Z}}\half
\left[ E\left((-2a+\alpha+2(\ell+2)\im
    z/y)\sqrt{y/\ell}\right) \right.  \non \\ 
 &&-\left. E\left(((2b-\beta)n-(2a-\alpha)
  m+2(2n+(\ell+2)m)\im\, z/y)\sqrt{y/J^2_{m,n}}\right) \right] \\ 
&&\times
w^{(\ell-2)(2b-\beta)+2(2a-\alpha)}\,q^{\frac{\ell}{4}(2b-\beta)^2+\frac{1}{2}(2b-\beta)(2a-\alpha)}\non
\end{eqnarray}
with $J_{m,n}^2=m(\ell m+2n)$. The completion for $f_{2,\beta
  C-\alpha f}$ follows directly from  $\widehat f_{2,\beta
  C-\alpha f}=\widehat A_{\ell,(\alpha,\beta)}+\widehat \vartheta
_{\alpha, \beta}^{m,n}$. The non-holomorphic term of the first line in Eq. (\ref{eq:wvarth}) is cancelled by the
non-holomorphic term of $\widehat
A_{\ell,(\alpha,\beta)}(z,\tau)$. Thus for the completion of $f_{2,\beta
  C-\alpha f}$ (and therefore also of $h_{2,\beta
  C-\alpha f}$) the non-holomorphic part of the second line in Eq. (\ref{eq:wvarth}) suffices. We define $
\mathcal{\widehat  Z}_r(\rho,z,\tau;S,J):=\sum_{\mu\in
  H^2(\Sigma_\ell,\mathbb{Z}/r\mathbb{Z})} \overline{\widehat
  h_{r,\mu}(z,\tau;S,J)}\,\Theta_{r,\mu}(\rho,\tau;S)$. 

\subsection{Holomorphic anomaly for rank 2}
\label{subsec:holanol} 
This subsection derives $D_r\mathcal{\widehat
  Z}_r(\rho,\tau;\Sigma_\ell,J)$ for $D_r=\partial_\tau+\frac{i}{4\pi
  r}\partial^2_{\rho_+}$ and $r=2$. Since
$D_r\Theta_{r,\mu}(\rho,\tau;\Sigma_\ell)=0$ for any $r$, it suffices to determine
$\partial_{\bar \tau}\widehat f_{r,c_1}(\tau;\Sigma_\ell,J)$. 
For a clear exposition, the generating functions are given in this
subsection in terms of $K_\ell$, $J$ etc. instead
of the explicit integers $\ell$, $m$ and $n$ etc. 

We determine first the completion $\widehat f_{2,c_1}(\tau;\Sigma_\ell,J)$
from the generating functions in the previous subsection. The result
follows from the following three steps: 
\begin{itemize}
\item[-] use Eq. (\ref{eq:fwtof}) after replacing the functions with their completions,
\item[-] use that $E(z)=2\int_0^z e^{-\pi
    u^2}du=\sgn(z)(1-\beta_{\frac{1}{2}}(z^2))$ with $z\in \mathbb{R}$
  and 
\be
\beta_{\nu}(x)=\int_x^\infty u^{-\nu}e^{-\pi u}du,\non
\ee
\item[-] and finally use 
\be
\beta_{\frac{3}{2}}(x)=2x^{-\frac{1}{2}}e^{-\pi x}-2\pi \beta_{\frac{1}{2}}(x).\non
\ee
\end{itemize}
One obtains:
\begin{eqnarray}
&&\widehat
f_{2,c_1}(\tau;\Sigma_\ell,J)=f_{2,c_1}(\tau;\Sigma_\ell,J)+ \non \\ 
&&\quad \sum_{\mathbf{c}\in -c_1\atop + H^2(\Sigma_\ell,
  2\mathbb{Z})}\left(\frac{K_{\ell}\cdot J\,|\mathbf{c}\cdot 
  J|}{8\pi J^2}\,\beta_{\frac{3}{2}}(\mathbf{c}_+^2\,y)
-\frac{1}{4} K_{\ell}\cdot \mathbf{c}_- \,\sgn(\mathbf{c}\cdot
J)\,\beta_{\frac{1}{2}}(\mathbf{c}_+^2\,y)\right)\,(-1)^{K_{\ell}\cdot
\mathbf{c}} q^{-\mathbf{c}^2},  \non 
\end{eqnarray}
where $\mathbf{c}_\pm$ are given by Eq. (\ref{eq:projections}). 
It is now straightforward to compute $\partial_{\bar \tau}\widehat
f_{r,c_1}(\tau;\Sigma_\ell,J)$:
\begin{eqnarray}
\partial_{\bar \tau}\widehat f_{r,c_1}(\tau;\Sigma_\ell,J)&=&\frac{i\,K_{\ell}\cdot
  J}{16\pi\,\sqrt{J^2}\, y^\frac{3}{2}}\, (-1)^{K_{\ell}\cdot
  c_1}\,\overline{\Theta_{2,-c_1-K_{\ell}}(0,\tau;\Sigma_\ell)} \non
\\ 
&&-\frac{i}{8\sqrt{y}} \sum_{\mathbf{c}\in -c_1\atop + H^2(\Sigma_\ell,
  2\mathbb{Z})} K_{\ell}\cdot
\mathbf{c}_-\,\frac{\mathbf{c}\cdot J}{\sqrt{J^2}}\,
\,(-1)^{K_{\ell}\cdot\mathbf{c}}\,
q^{-\mathbf{c}_-^2/4} \bar q^{\mathbf{c}_+^2/4} .\non 
\end{eqnarray}
After combining this result with
$\Theta_{2,c_1}(\rho,\tau;\Sigma_\ell)$ as in (\ref{eq:thetadecomp})
and manipulation of the lattice sums, one obtains for  $D_2\mathcal{\widehat Z}_2(\rho,\tau;\Sigma_\ell,J)$:
\begin{eqnarray}
\label{eq:holanomaly}
D_2 \mathcal{\widehat Z}_2(\rho,\tau;\Sigma_\ell,J)&=&\frac{-i\,K_{\ell}\cdot
  J}{16\pi\,\sqrt{J^2}\,
  y^\frac{3}{2}}\,\mathcal{Z}_1(\rho,\tau;\Sigma_\ell,J)^2\\
&&+\frac{i}{8\sqrt{y}}\, \overline{h_{1,0}(\tau;\Sigma_\ell)}^2\,\sum_{c_1\in H^2(\Sigma_\ell,\mathbb{Z}/2\mathbb{Z})}\Upsilon_{c_1}(\tau,\Sigma_\ell) \,\Theta_{2,c_1}(\rho,\tau;\Sigma_\ell), \non
\end{eqnarray}
where\footnote{A similar function appeared in Ref. \cite{Manschot:2009ia}.}
\be
\Upsilon_{c_1}(\tau,\Sigma_\ell)=\sum_{\mathbf{c}\in -c_1\atop + H^2(\Sigma_\ell,
  2\mathbb{Z})} K_{\ell}\cdot
\mathbf{c}_-\,\frac{\mathbf{c}\cdot J}{\sqrt{J^2}}\,
\,(-1)^{K_{\ell}\cdot\mathbf{c}}\,
q^{\mathbf{c}_+^2/4} \bar q^{-\mathbf{c}_-^2/4}. \non  
\ee
Interestingly, Eq. (\ref{eq:holanomaly}) differs from the conjectured
form of the anomaly \cite{Vafa:1994tf, Minahan:1998vr, Alim:2010cf}. The first line has the
expected factorized form, which is attributed to reducible
connections or polystable sheaves \cite{Vafa:1994tf} or multiple M5-branes \cite{Minahan:1998vr}. However, the
novel second line does not factorize and is less easily interpreted.  It does vanish for
special values of $J$, in particular for $J=-K_\ell$
since then $K_\ell\cdot \mathbf{c}_-=0$. But for $\ell\geq 2$,
$K_\ell$ lies outside $C(S)$ and is thus not a permissible choice
for $J$. Viewing the surface as part of a local Calabi-Yau 3-fold
geometry, $J=-K_S$ corresponds to the attractor point from
the point of view of supergravity \cite{Manschot:2009ia}. It is
therefore rather interesting that $\mathcal{\widehat
  Z}(\rho,\tau;\Sigma_\ell,J)$ simplifies at this point. 

The function $\Upsilon_{c_1}(\tau,\Sigma_\ell)$ vanishes also for 
 $\ell=1$ and $J=C+ f$ \cite{Bringmann:2010sd}, which is not equal to
 $-K_1$. For this choice, the blow-up formula gives 
 the generating function for $\mathbb{P}^2$, where $J=-K_{\bP^2}$ is
 satisfied automatically. It is thus in agreement with these
 examples to conjecture that generically for a K\"ahler surface $S$,
$D_2 \mathcal{\widehat Z}_2(\rho,\tau;S,J)=\frac{-i\,\sqrt{K_{S}^2}}{16\pi\,
  y^\frac{3}{2}}\,\mathcal{Z}_1(\rho,\tau;S,-K_s)^2$ if $K_S\in
C(S)$. Of course, a more intrinsic explanation based on gauge theory
or algebraic geometry is desirable.

%
%
%
 
\subsection{Rank 3} 
\label{subsec:rank3}
This subsection presents the generating functions  $h_{3,\beta C-\alpha f}(z,\tau;\Sigma_\ell,J)$ with $\beta\neq 0
\mod 3$. This condition on $\beta$ ensures that $h_{3,\beta C-\alpha
  f}(z,\tau;\Sigma_\ell,J)=0$ for $J=J_{0,1}$ analogously to $r=2$. The computation of $h_{3,\beta C-\alpha
  f}(z,\tau;\Sigma_\ell,J)$
therefore reduces again to application of the wall-crossing
formula.  This is for $r=3$ more complicated than for $r=2$ since:
\begin{itemize}
\item[-] the functions $h_{2,c_1}(z,\tau;\Sigma_\ell,J)$ do themselves
  depend on $J$, and need to be determined sufficiently close to the appropriate wall. 
\item[-] the total charge $\Gamma$ can be of a sum of 3 charges
  $\sum_{i=1}^3 \Gamma_i$ such that at a wall $W$ the slopes of these
  three constituents might be equal. This in particular happens for
  ``semi-primitive wall-crossing'' where $\Gamma(F)=2\Gamma_1+\Gamma_2$.
\end{itemize}
Nevertheless, the wall-crossing formulas \cite{Kontsevich:2008, 
  Joyce:2008} imply a relatively simple form for the generating functions \cite{ Manschot:2010xp,arXiv:1009.1775}. 
One obtains for $\ell\geq 1$:
\begin{eqnarray}
\label{eq:r3w}
 f_{3,\beta C-\alpha f}(z,\tau;\Sigma_\ell, J_{m,n})&=&-\sum_{a,b\in \mathbb{Z}}\half (\, \sgn((3b-2\beta)n-(3a-2\alpha )m)-\sgn(3b-2\beta)
\,)\non \\
&&\times \left(w^{(\ell-2)(3b-2\beta)+2(3a-2\alpha)}-w^{
    -(\ell-2)(3b-2\beta)-2(3a-2\alpha)}
\right)\\
&&\times\,
q^{\frac{\ell}{12}(3b-2\beta)^2+\frac{1}{6}(3b-2\beta)(3a-2\alpha)}\non \\
&& \times\,
f_{2,bC-af}(z,\tau;\Sigma_\ell,J_{|3b-2\beta|,|3a-2\alpha|}),\non 
\end{eqnarray}
for $\beta=1,2 \mod 3$ and $\alpha\in \bZ$. Writing out the lattice
sums in Eq. (\ref{eq:r3w}), one finds a novel indefinite theta
function. It has signature $(2,2)$ and the condition which determines
whether or not a lattice point contributes depends quadratically on the lattice
vector, whereas previously described indefinite theta functions
have signature $(n,1)$ and the condition depends linearly on the
lattice vector \cite{Gottsche:1996, Zwegers:2000}.
A detailed discussion of the (mock) modular properties of $h_{3,c_1}(z,\tau;\Sigma_\ell,J)$
will appear in a future article \cite{Bringmann:2011}. 

Tables \ref{tab:betti1}-\ref{tab:betti3} list Betti numbers for
$c_1=-C-\alpha f$ with $\alpha=1,2,3$ and $\ell=1$, which are in
agreement with the expected dimension (\ref{eq:dimension}).
One can relate the Betti numbers for
$c_1=-2C-\alpha f$ to these by using $h_{r,c_1}=h_{r,-c_1}$, and
$h_{r,c_1+\mathbf{k}}=h_{r,c_1}$ for $\mathbf{k}\in
H_2(S,r\mathbb{Z})$. With a little more work, one can verify that 
$h_{3,c_1}(z,\tau;\Sigma_1,J_{1,0})$ satisfies the relations
implied by the blow-up formula (\ref{eq:blowup}).



\begin{table}[h!]
\begin{tabular}{lrrrrrrrrrrrrrr}
$c_2$ & $b_0$ & $b_2$ & $b_4$ & $b_6$ & $b_8$ & $b_{10}$ & $b_{12}$ & $b_{14}$
& $b_{16}$ & $b_{18}$ & $b_{20}$ & $b_{22}$ & $b_{24}$ &  $\chi$ \\
\hline
2 & 1 & 2 & 4 & 4  &  & & & & & & & & & 18  \\
3 & 1 & 3 & 9 & 20 & 37 & 53 & 59 & & & & & & & 305 \\
4 & 1 & 3 & 10 & 25 & 59  & 119 & 218 & 338 & 450 & 490 &&&& 2936 \\
5 & 1 & 3 & 10 & 26 & 64 & 141 & 294 &  562 & 997 & 1602 & 2301 & 2886 & 3117 & 20891 \\  
&
\end{tabular}
\caption{The Betti numbers $b_n$ (with $n\leq
  \dim_\mathbb{C} \mathcal{M}$) and the Euler numbers $\chi$ of the moduli spaces of stable sheaves
  on $\Sigma_1$ with $r=3$, $c_1=-C$, and $2\leq c_2\leq 6$ for $J=(1,\varepsilon)$.}  
\label{tab:betti1}
\end{table}

\begin{table}[h!]
\begin{tabular}{lrrrrrrrrrrrrrrr}
$c_2$ & $b_0$ & $b_2$ & $b_4$ & $b_6$ & $b_8$ & $b_{10}$ & $b_{12}$ & $b_{14}$
& $b_{16}$ & $b_{18}$ & $b_{20}$ & $b_{22}$ & $b_{24}$ & $b_{26}$ & $\chi$ \\
\hline
2 & 1 & 1 & & & & & & & & & & & & & 3 \\
3 & 1 & 3 & 8 & 14 & 17 & & & & & & & & & & 69  \\
4 & 1 & 3 & 10 & 24 & 53  & 93 & 136 & 152 & &&&&& & 792\\
5 & 1 & 3 & 10 & 26 & 63 & 135 & 268 &  470 & 725 & 950 & 1043 & && & 6345\\
6 & 1 & 3 & 10 & 26 & 65 & 145 & 310 & 612 & 1144 & 1970 & 3113 & 4391 &
5462 & 5873 & 40377 \\
&
\end{tabular}
\caption{The Betti numbers $b_n$ (with $n\leq
  \dim_\mathbb{C} \mathcal{M}$) and the Euler numbers $\chi$ of the moduli spaces of stable sheaves
  on $\Sigma_1$ with $r=3$, $c_1=-C-f$, and $2\leq c_2\leq 6$.}  
\label{tab:betti2}
\end{table}

\begin{table}[h!]
\begin{tabular}{lrrrrrrrrrrrrrrr}
$c_2$ & $b_0$ & $b_2$ & $b_4$ & $b_6$ & $b_8$ & $b_{10}$ & $b_{12}$ & $b_{14}$
& $b_{16}$ & $b_{18}$ & $b_{20}$ & $b_{22}$  & $\chi$ \\
\hline
3 & 1 & 2 & 3 & & & & & & & & & &  9 \\
4 & 1 & 3 & 9 & 19 & 31 & 36 & & & & & & &  162  \\
5 & 1 & 3 & 10 & 25 & 58  & 113 & 192 & 264 & 297 &&&& 1629 \\
6 & 1 & 3 & 10 & 26 & 64 & 140 & 288 &  536 & 907 & 1348 & 1733 & 1885 &11997\\
&
\end{tabular}
\caption{The Betti numbers $b_n$ (with $n\leq
  \dim_\mathbb{C} \mathcal{M}$) and the Euler numbers $\chi$ of the moduli spaces of stable sheaves
  on $\Sigma_1$ with $r=3$, $c_1=-C-2f$, and $3\leq c_2\leq 6$.}  
\label{tab:betti3}
\end{table}

\section*{Acknowledgements}
\noi I would like to thank L. G\"ottsche, B. Haghighat, H. Nakajima and
K. Yoshioka for helpful and inspiring discussions, and the LPTHE and IHES
for hospitality. This work is  partially supported  by ANR grant
BLAN06-3-137168.

\appendix
\section{Modular functions}
\label{app:modfunctions}

Define $q:=e^{2\pi i \tau}$, $w:= e^{2\pi i z}$, with $\tau \in \mathbb{H}$ and $z\in \mathbb{C}$. 
The Dedekind eta and Jacobi theta functions are defined by: 
\begin{eqnarray}
\label{eq:etatheta}
&&\eta(\tau)\quad \,\,:=q^{\frac{1}{24}}\prod_{n=1}^\infty (1-q^n),\non\\
&&\theta_1(z,\tau):=i\sum_{r\in \mathbb{Z}+\frac{1}{2}} (-1)^{r-\frac{1}{2}} q^{\frac{r^2}{2}}w^{r},\\
&&\theta_2(z,\tau):=\sum_{r\in \mathbb{Z}+\frac{1}{2}} q^{r^2/2}w^{r},\non\\
&&\theta_3(z,\tau):=\sum_{n\in\mathbb{Z}}q^{n^2/2}w^n\non.
\end{eqnarray}

The Appell function at level $\ell$ is defined by:
\be
\label{eq:appell}
A_\ell(u,v,\tau)=a^{\ell/2}\sum_{n\in
  \mathbb{Z}}\frac{(-1)^{\ell n}q^{\ell n(n+1)/2}b^n}{1-aq^n}, 
\ee
with $a=e^{2\pi i u}$ and $b=e^{2\pi i v}$. In order to give the
completion $\widehat A_\ell(u,v,\tau) $, define
\begin{eqnarray}
R(u,\tau)&=&\sum_{r \in \mathbb{Z}+\frac{1}{2}} \left(\, \sgn(r)-E\left(
    (r+\im\, u/y)\sqrt{2 y}\right)\,\right)\non\\
&& \times (-1)^{r-\frac{1}{2}}a^{-r}q^{-r^2/2}, \non
\end{eqnarray}
with $E(x)=2\int_0^x e^{-\pi u^2}du$. The completion $\widehat
A_\ell(u,v,\tau)$ is then given by \cite{Zwegers:2010} 
\begin{eqnarray} 
\label{eq:appellcom}
\widehat A_\ell(u,v,\tau)&=&A_\ell(u,v,\tau)+\frac{i}{2}\sum_{k=0}^{\ell-1}\,a^k\,\theta_1(v+k\tau+(\ell-1)/2,\ell
\tau)\\
&&\times R(\ell u-v-k\tau-(\ell-1)/2,\ell \tau),
\non 
\end{eqnarray}
and transforms as a multivariable Jacobi form of weight 1 and index
$\frac{1}{2} \left(\begin{array}{cc} -\ell & 1 \\ 1 & 0 \end{array}\right)$.
The Appell function for $\ell=1$ is related to the Lerch-Appell
function: $\mu(u,v,\tau)=A_1(u,v,\tau)/\theta_1(v)$, which satisfies the 
quasi-periodicity property \cite{Zwegers:2000}:
\be
\label{eq:quasiperiod}
\mu(u+z,v+z,\tau)-\mu(u,v,\tau)=\frac{\eta(\tau)^3\,\theta_1(u+v+z,\tau)\,\theta_1(z,\tau)}{\theta(u,\tau)\,\theta(v,\tau)\,\theta(u+z,\tau)\,\theta(v+z,\tau)},
\ee
for $u,v,u+z,v+z\notin \mathbb{Z}\tau+\mathbb{Z}$. 

\providecommand{\href}[2]{#2}\begingroup\raggedright\endgroup

\end{document}